\documentclass[pra,amssymb,amsfonts,amsmath,superscriptaddress,showpacs]{revtex4}

\newcommand{\tr}{\mbox{Tr}}
\newcommand{\bra}[1]{\ensuremath{\langle #1 |}}
\newcommand{\ket}[1]{\ensuremath{| #1 \rangle}}
\newcommand{\bk}[2]{\ensuremath{\langle #1 | #2 \rangle}}
\newcommand{\kb}[2]{\ensuremath{| #1 \rangle\!\langle #2 |}}

\newtheorem{re}{Remark}
\newtheorem{theo}{Theorem}

\newtheorem{lem}{Lemma}

\begin{document}

\title{`Classical' quantum states}
\author{Marek Ku\'s}
\affiliation{Center for Theoretical Physics, Polish Academy of Sciences,
Aleja Lotnik{\'o}w 32/44, 02-668 Warszawa, Poland}
\author{Ingemar Bengtsson}
\affiliation{\it Stockholm University, AlbaNova, Fysikum, 106 91 Stockholm,
Sweden.}

\begin{abstract}
We show that several classes of mixed quantum states in finite-dimensional
Hilbert spaces which can be characterized as being, in some respect, 'most
classical' can be described and analyzed in a unified way. Among the states
we consider are separable states of distinguishable particles, uncorrelated
states of indistinguishable fermions and bosons, as well as mixed spin states
decomposable into probabilistic mixtures of pure coherent states. The latter
were the subject of the recent paper by Giraud et.\ al.\ \cite{giraud08}, who
showed that in the lowest-dimensional, nontrivial case of spin 1, each such
state can be decomposed into a mixture of eight pure states. Using our method
we prove that in fact four pure states always suffice.
\end{abstract}

\date{\today}
\maketitle

\section{Introduction}
The notion of being the `most classical' among all quantum states of a given
system is admittedly vague, and strongly dependent on the particular quantum
features we would like to approximate on the classical level --
`nonclassical' properties of quantum systems usually encompass a variety of
phenomena.

In the situation that we will be concerned with here, there is a Lie group of
preferred observables acting on the system. It may be the group $SU(N)\times
SU(M)$ of local unitaries acting on a composite system, or the $SU(2)$ group
one can conveniently realize using linear optics. Then the group action
divides the set of states into disjoint orbits, and there will be a special
orbit which is in itself a symplectic and indeed a K\"ahler manifold. It can
serve as a classical phase space, and the states in this orbit are our `most classical'
states. They are separable states in the first example, and coherent spin
states in the second. In general they are the generalized coherent states
\cite{perelomov86} defined as the orbit in the representation space of the
highest weight vector for a particular irreducible representation of the
symmetry group of the system. Generalized coherent states also minimize an
uncertainty relation \cite{delbourgo77a}, or more precisely they minimize the
total variance of a state $\psi$,
\begin{equation}\label{unc}
\delta(\psi)=\sum_i\left(\bra{\psi}X_i^2\ket{\psi}-\bra{\psi}X_i\ket{\psi}^2\right),
\end{equation}
where the $\{X_i\}$ form an orthonormal basis in the Lie algebra of the
relevant group. Note that this is independent of the choice of the basis
$\{X_i\}$. It is pleasing that both definitions of `most classical' coincide,
and moreover, that they coincide with the view that separable states of a
composite system are the most classical ones, in the sense that they lack the
quantum correlations among subsystems responsible for any EPR-like phenomena.

In the following we call (for shortness) those states `classical' which are
`most classical' in the above senses. At the other end of the spectrum we can
identify the states in that special orbit on which the symplectic form
vanishes as being the `most non-classical' states. The reason why there is
such an orbit can be seen by complexifying the group; for composite system
this coincides with the maximally entangled states as conventionally
understood \cite{klyachko08}.

Investigations concerning role of pure (generalized) coherent states in
exhibiting links between quantum and classical level have a long history,
especially for such problems as quantization and classical limits (see eg.
\cite{klauder:85}). The connections between coherent states and entanglement
theory was recently pointed by Klyachko \cite{klyachko02,klyachko08}
\textbf{}(see also \cite{barnum03,barnum04,barnum05}).

The notion of classicality can be extended to mixed states. On a formal level
we do it by defining `classical' mixed states as those which can be
decomposed in a probabilistic mixture (i.e.\ a convex linear combination) of
pure classical states (here and in the following we identify pure states,
i.e.\ vectors in a Hilbert space, or rather its projectivisation, with
one-dimensional projection on their directions). For composite systems it
leads to the popular definition of separable (or equivalently non-entangled)
states, which can be also characterized on the operational level as those
which can be obtained from an initially uncorrelated state of the whole
system by local operations, i.e.\ quantum operations performed only on
subsystems, accompanied by possible exchange of classical information between
them (e.g.\ about results of local measurement on which further local
operations can be conditioned) \cite{bengtsson06}. The general practical
criteria useful for unambiguous discrimination of mixed separable states
among all states of a composite quantum system are not known, but in
low-dimensional cases there are several methods (e.g.\ based on determining
the so called concurrence of a state) to achieve the goal.

Equating absence of genuine quantum correlations with the property of being a
product state (simple tensor) lacks sense in the case of indistinguishable
particles. Indistinguishability forces (anti)symmetrization of wavefunctions,
in other words states are vectors not in tensor products of Hilbert spaces
but in their (anti)symmetrizations. With the exception bosons occupying the
same state, no wave function has the form of a simple tensor. It is
reasonable to call `nonclassical' quantum correlations which go beyond those
stemming solely from (anti)symmetrization, and consequently states which lack
such correlations are treated as quantum uncorrelated or the best candidates for
being `most classical'.  Methods of investigating and characterizing such
states were developed in \cite{sckll01} and \cite{eckert02}.

For spin-coherent states, when the group is $SU(2)$, the classical mixed
states---as is customary in quantum optics---are characterized as having a
positive $P$-representation. The definition is equivalent to decomposability
of a state into a probabilistic mixture of pure coherent states. For the
lowest-dimensional nontrivial case (spin 1) Giraud et.\ al.\ \cite{giraud08}
give a necessary and sufficient condition for classicality of the state.
Their proof of the correctness of the criterion is constructive --- it gives
explicitly a decomposition into eight coherent states.

The observation that all above mentioned cases are instances of the
general construction of coherent states allows to analyze them in a unified
manner. One may thus use methods known from the theory of coherent states to
the separability problem, or \textit{vice versa}, apply techniques from
entanglement theory to exhibit some new features of generalized coherent
states. We will mainly follow the latter path --- a concrete aim is to refine one of
the results of \cite{giraud08} concerning the cardinality of the optimal
decomposition of a mixed spin-coherent state into a mixture of pure coherent
states. Along the way we present a unified treatment of correlations
in two-component quantum systems in terms of bilinear observables. We start
by recalling the characterization of entanglement by the
(pre)concurrence, defined originally by Wootters \cite{wootters98} and
elaborated further by Uhlmann \cite{uhlmann00}. In this approach one
quantifies the entanglement of a (pure) state by the expectation value of
some antilinear operator and extends the measure to mixed states by the
`convex roof' construction, i.e.\ by decomposing a mixed state into a
probabilistic mixture of pure states and minimizing over all possible
decompositions the sum of concurrences of the components. As observed by
Uhlmann \cite{uhlmann00}, antilinear operators `are intrinsically nonlocal',
and as such are good candidates for a tool probing nonlocal properties
like entanglement. On the other hand, they do not correspond to physical
observables, hence their expectation values can not be directly measured.
Elsewhere \cite{mckb05,mkb04} we argued that it might be more convenient to
use bilinear Hermitian operators to quantify entanglement --- such an
approach is efficient in higher dimensions and provides measures of
entanglement accessible directly in experiments \cite{walborn06}. Remarkably
the same approach enables also a unified treatment of the different facets of
`classicality' enumerated above.

\section{Generalized concurrences}
We start with an elucidation of links between generalized concurrences
defined by Uhlmann and bilinear operators. As already advertised in the
previous section, both provide tools for characterizing separability and
entanglement as we will show in Section~\ref{sec:examples}.

In the following $\mathcal{H}$ will denote a finite-dimensional complex
Hilbert space and $\{\ket{e_i}\}_{i=1}^N$ is an orthogonal basis in
$\mathcal{H}$. Let $\psi\in\mathcal{H}$ and $C_\Theta(\psi)$ ---the
generalized concurrence in the sense of Uhlmann \cite{uhlmann00}---be defined
\emph{via} an antiunitary operator $\Theta$, i.e.
\begin{equation}\label{uhlmann}
C_\Theta(\psi)=|\bra{\psi}\Theta\ket{\psi|}|.
\end{equation}
Each antiunitary $\Theta$ can be written as $\Theta=T_\Theta\,K$, where
$T_\Theta$ is unitary and $K$ is the complex conjugation in some fixed basis
(say the chosen one $\{\ket{e_i}\}_{i=1}^N$), i.e.
\begin{equation}
C_\Theta(\psi)=|\bra{\psi}T_\Theta\ket{\psi^\ast}|.
\end{equation}
Let us define $A\in End(\mathcal{H}\otimes\mathcal{H})$,
\begin{equation}\label{jam}
A=\big(\mathbb{I}\otimes \Lambda\big)\kb{\Phi}{\Phi},
\end{equation}
where $\ket{\Phi}$ is the maximally entangled state in
$\mathcal{H}\otimes\mathcal{H}$,
\begin{equation}\label{maxent}
\ket{\Phi}=\sum_{i=1}^N\ket{e_i}\otimes\ket{e_i},
\end{equation}
and for an arbitrary $\rho\in End(\mathcal{H})$
\begin{equation}
\Lambda(\rho)=T_\Theta\rho T_\Theta^\dagger,
\end{equation}
i.e.\ $A$ is the image under the Jamio{\l}kowski isomorphism of $\Lambda$
which, in turn, is a completely-positive unitary map on
$End(\mathcal{H}\otimes\mathcal{H})$. Since $\Lambda$ is completely-positive,
$A$ is non-negatively definite \cite{jamiolkowski72,choi75,bengtsson06}.

The announced connection between the bilinear operator $A$ and the
generalized concurrence $C_\Theta$ is given by the following
\begin{lem}\label{lem:A-conc}
\begin{equation}\label{bilinear}
C_\Theta(\psi)=\bra{\psi\otimes\psi} A\ket{\psi\otimes\psi}^{1/2}.
\end{equation}
\end{lem}

\underline{\bf Proof}

From the definition of $\ket{\Phi}$ and $\Lambda$
\begin{eqnarray}\label{Ajam}
A=\big(\mathbb{I}\otimes
\Lambda\big)\sum_{i,j}\kb{e_i}{e_j}\otimes\kb{e_i}{e_j}=
\sum_{i,j}\kb{e_i}{e_j}\otimes
T_\Theta\kb{e_i}{e_j}T_\Theta^\dagger.
\end{eqnarray}
Hence, for $\ket{\psi}=\sum_k c_k\ket{e_k}$,
\begin{eqnarray}\label{psiApsi}
\bra{\psi\otimes\psi}A\ket{\psi\otimes\psi}=\sum_{ijklmn}c_k^\ast c_l^\ast
c_m c_n
\bk{e_k}{e_i}\bk{e_j}{e_m}\bra{e_l}T_\Theta\kb{e_i}{e_j}T_\Theta^\dagger
\ket{e_n} 
=\sum_{ijklmn}c_k^\ast c_l^\ast c_m c_n
\delta_{ki}\delta_{jm}\bra{e_l}T_\Theta\kb{e_i}{e_j}T_\Theta^\dagger
\ket{e_n}\nonumber \\
=\sum_{ijln}c_i^\ast c_l^\ast c_j c_n
\bra{e_l}T_\Theta\kb{e_i}{e_j}T_\Theta^\dagger
\ket{e_n}=\bra{\psi}T_\Theta\ket{\psi^\ast}\bra{\psi^\ast}
T_\Theta^\dagger\ket{\psi}
=\bra{\psi}T_\Theta\ket{\psi^\ast}\bra{\psi}T_\Theta\ket{\psi^\ast}^\ast
=|\bra{\psi}T_\Theta\ket{\psi^\ast}|^2=C_\Theta(\psi)^2.
\end{eqnarray}
$\Box$
\begin{re} We have
\begin{equation}\label{normA}
\mathrm{Tr}_1 A=\mathbb{I},
\end{equation}
and
\begin{equation}\label{normA1}
\mathrm{Tr}_2 A=\mathbb{I},
\end{equation}
where ${\tr}_{1,2}$ denotes the partial trace over the first or the second
copy of $\mathcal{H}$.  Indeed, from the unitarity of $T_\Theta$,
\begin{eqnarray}
\mathrm{Tr}_1 A=\sum_{ijk}\bk{e_k}{e_i}\bk{e_j}{e_k}
T_\Theta\kb{e_i}{e_j}T_\Theta^\dagger=
\sum_{ijk}\delta_{ki}\delta_{jk}T_\Theta\kb{e_i}{e_j}T_\Theta^\dagger
=T_\Theta\Big(\sum_{i}\kb{e_i}{e_i}\Big)T_\Theta^\dagger
=T_\Theta T_\Theta^\dagger=\mathbb{I},
\end{eqnarray}
and
\begin{eqnarray}
\mathrm{Tr}_2 A&=&\sum_{ijk}\kb{e_i}{e_j}\bra{e_k}
T_\Theta\kb{e_i}{e_j}T_\Theta^\dagger\ket{e_k}=
\sum_{ijk}\kb{e_i}{e_j}\big(T_\Theta^\dagger\big)_{jk}\big(T_\Theta\big)_{ki}
=\sum_{ij}\kb{e_i}{e_j}\big(T_\Theta^\dagger T_\Theta\big)_{ji}
\nonumber \\
&=&\sum_{ij}\kb{e_i}{e_j}\delta_{ji}=\sum_{i}\kb{e_i}{e_i}=\mathbb{I}.
\end{eqnarray}
\end{re}

Observe now that we may proceed in the opposite way, i.e.\ define the
generalized concurrence as in (\ref{bilinear}),
\begin{equation}\label{CA}
C_A(\psi)=\bra{\psi\otimes\psi} A\ket{\psi\otimes\psi}^{1/2},
\end{equation}
where $A$ is some non-negatively defined linear operator on
$\mathcal{H}\otimes\mathcal{H}$ with the property (\ref{normA}). Then we
define $\Lambda\in End\big(End(\mathcal{H})\big)$ as the image of $A$ under
the inverse of the Jamio{\l}kowski isomorphism, i.e.,
\begin{equation}\label{iJ}
 \Lambda(\rho)=\mathrm{tr}_1\Big((\rho^t\otimes\mathbb{I}) A\Big),
\end{equation}
for $\rho\in End(\mathcal{H})$. Observe that from (\ref{normA}) and
(\ref{iJ}),
\begin{equation}\label{unimodal}
\Lambda(\mathbb{I})=\mathrm{tr}_1(A)=\mathbb{I}.
\end{equation}

Since $A$ is non-negatively defined, $\Lambda$ is completely positive, hence
it has a Kraus decomposition \cite{choi75,bengtsson06},
\begin{equation}\label{kraus}
\Lambda(\rho)=\sum_{\alpha=1}^sT_\alpha\rho T_\alpha^\dagger.
\end{equation}

If $A$ is such that $s=1$, then $\Lambda\rho=T_1\rho T_1^\dagger$ and from
(\ref{unimodal}) $T_1T_1^\dagger=\mathbb{I}$, i.e. $T_1$ is unitary. Then we
have
\begin{equation}
C_A(\psi)=|\bra{\psi}T_1\ket{\psi^\ast}|=C_\Theta(\psi),
\end{equation}
where $\Theta=T_1K$. To prove the last statement it is enough to check that
(\ref{iJ}) gives indeed the inverse of the Jamio{\l}kowski isomorphism
(\ref{jam}) and perform the calculations in the proof of
Lemma~\ref{lem:A-conc} in the reverse direction.

It seems to be more appropriate to take (\ref{CA}) with the condition
(\ref{normA}) as the definition of the generalized concurrence (see
\cite{mckb05}). Generically the non-negative definite matrix $A$ has the
spectral decomposition
\begin{equation}
A=\sum_{\alpha=1}^s \nu_\alpha \kb{v_\alpha}{v_\alpha}=\sum_{\alpha=1}^s
\kb{w_\alpha}{w_\alpha},
\end{equation}
with more then one nonvanishing eigenvalues $\nu_\alpha$. Here
$\ket{w_\alpha} =\sqrt{\nu_\alpha}\,\ket{v_\alpha}$ are the subnormalized
eigenvectors of $A$. It means that $\Lambda$ defined by (\ref{iJ}) is given
by (\ref{kraus}) with
\begin{equation}\label{Ta}
T_\alpha=\Big(\bra{\Phi}\otimes\mathbb{I}\Big)\Big(\mathbb{I}
\otimes\ket{w_\alpha}\Big).
\end{equation}
Only exceptionally (eg.\ in low-dimensional cases) $s=1$ and we can write
$C_A$ in terms of an antiunitary operator $\Theta$.

Performing the same calculations as in (\ref{psiApsi}) for the general case
(\ref{kraus}) we obtain:
\begin{equation}\label{psiApsigen}
\bra{\psi\otimes\psi}A\ket{\psi\otimes\psi}=\sum_{\alpha=1}^s
\bra{\psi}T_\alpha\ket{\psi^\ast}\bra{\psi^\ast}
T_\alpha^\dagger\ket{\psi}=\sum_{\alpha=1}^s|\bra{\psi}T_\alpha\ket{\psi^\ast}|^2,
\end{equation}
and by polarization,
\begin{equation}
 \bra{\psi_2\otimes\psi_4}A\ket{\psi_1\otimes\psi_3}=\sum_{\alpha=1}^s
\bra{\psi_2}T_\alpha\ket{\psi_4^\ast}\bra{\psi_1^\ast}
T_\alpha^\dagger\ket{\psi_3}.
\end{equation}

\section{Mixed states}

The concept of concurrence can be extended to mixed states.  The idea is
based on the following observation \cite{gkm06} (see also
\cite{uhlmann00,vidal00}).

Let $E$ be the set of all extreme points of a compact convex set $K$ in a
finite dimensional real vector space $V$. For every non-negative function $f
: E \rightarrow \mathbb{R}_+$ we may define its extension $f_K : K
\rightarrow \mathbb{R}_+$ by
\begin{equation}\label{roof}
f_K(x)=\inf_{x=\sum p_i x_i}\sum p_if( x_i)
\end{equation}
where the infimum is taken with respect to all expressions of $x$ in the form
of convex combinations of points $x_i$ from $E$. Let now $E_0$ be a compact
subset of $E$ with the convex hull $K_0 = conv(E_0)\subset K$. If $f$ is
continuous and vanishes exactly on $E_0$, then the function $f_K$ is convex
on $K$ and vanishes exactly on $K_0$. In our cases $K$ is the set of all
states, $E$ --- the set of pure states, and $E_0$ --- the set of pure
`classical' states. Observe that due to homogeneity of the generalized
concurrence, $C_A(\alpha\psi)=|\alpha|^2C_A(\psi)$, we can consider only the
decompositions into sums of rank-one operators, defining thus for an
arbitrary state $\rho$,
\begin{equation}\label{mixedC}
C_A(\rho)=\min_{\{\phi_k\}}\sum_{k=1}^K C_A(\phi_k),
\end{equation}
where the minimum is taken over all decompositions of $\rho$ into a sum of
rank-one operators,
\begin{equation}\label{gendecomp}
\rho=\sum_{k=1}^K\kb{\phi_k}{\phi_k}.
\end{equation}
We took advantage of the finite dimensionality of the Hilbert space and
substituted the minimum for the the infimum operation \cite{uhlmann97} --- in
the finite-dimensional case there exists always an optimal decomposition,
i.e.\ one which minimizes $C_A(\rho)$.

A particular example of (\ref{gendecomp}) can be obtained from the spectral
decomposition of $\rho$,
\begin{equation}\label{spectralrho}
\rho=\sum_{i=1}^r p_i\kb{\eta_i}{\eta_i},\quad
\rho\ket{\eta_i}=p_i\ket{\eta_i},
\quad \bk{\eta_i}{\eta_j}=\delta_{ij}, \quad r=\mathrm{rank}\rho,
\end{equation}
by subnormalizing the eigenvectors,
\begin{equation}\label{spectralrho1}
\ket{\xi_i}=\sqrt{p_i}\ket{\eta_i},\quad \rho=\sum_{i=1}^r\kb{\xi_i}{\xi_i}.
\end{equation}
Any other decomposition (\ref{gendecomp}) can be obtained from
(\ref{spectralrho1}) with the help of a partial isometry \cite{hughston93},
\begin{equation}\label{pisom}
 \ket{\phi_k}=\sum_{j=1}^r V_{kj}\ket{\xi_j},\quad k=1,\ldots,K;
\quad V^\dagger V=I.
\end{equation}
Hence,
\begin{eqnarray}\label{carho-exp}
C_A(\rho)&=&\min\sum_k C_A(\phi_k)=\min\sum_k\bra{\phi_k\otimes\phi_k}A
\ket{\phi_k\otimes\phi_k}^{1/2}
\nonumber \\
&=&\min\sum_k\big(V_{ki}^\ast V_{kj}^\ast V_{kl} V_{km}
\bra{\xi_i\otimes\xi_j}A\ket{\xi_l\otimes\phi_m}\big)^{1/2}
\nonumber \\
&=&\min\sum_k\Big(V_{ki}^\ast V_{kj}^\ast V_{kl} V_{km}
\sum_\alpha\bra{\xi_i}T_\alpha\ket{\xi_j^\ast}\bra{\xi_l^\ast}
T_\alpha^\dagger\ket{\xi_m}\Big)^{1/2}
\nonumber \\
&=&\min\sum_k\Big(\sum_\alpha(V^\ast\tau_\alpha V^\dagger)_{kk}
(V\tau_\alpha^\ast V^T)_{kk}\Big)^{1/2}
=\min\sum_k\Big(\sum_\alpha|(V^\ast\tau_\alpha
V^\dagger)_{kk}|^2\Big)^{1/2},
\end{eqnarray}
where $\tau_\alpha$ are $r\times r$ matrices,
\begin{equation}\label{tau}
(\tau_\alpha)_{ij}=\bra{\xi_i}T_\alpha\ket{\xi_j^\ast},
\end{equation}
and the minimum is taken over all partial isometries $V$. This can be a
starting point for various estimations of $C_A(\rho)$ in the spirit of
\cite{mckb05} and \cite{mkb04}.

In the case when $s=1$ (so the generalized concurrence for pure states can be
expressed in terms of an antiunitary operator (\ref{uhlmann})), the
expression (\ref{carho-exp}) reduces to
\begin{equation}
 C_A(\rho)=\min\sum_{k}|(V^\ast\tau V^\dagger)_{kk}|, \quad
\tau_{ij}=\bra{\xi_i}T_1\ket{\xi_j^\ast}.
\end{equation}
If $T_1$ is symmetric, $T_1=T_1^T$, (as it happens in all the cases we consider -
see below), then $\Theta=T_1K$ is an antiunitary conjugation, i.e.\
$\Theta^2=1$, and $C_A(\rho)$ can be explicitly calculated,
\begin{equation}\label{mainformula}
C_A(\rho)=\max\Big\{0,\mu_1-\sum_{j=2}^r\mu_j\Big\},
\end{equation}
where $\mu_i$ are singular values of $\tau$ in decreasing order. A short
calculation shows that $\mu_i$ are equal to the square roots of the
eigenvalues of $\rho\tilde{\rho}$, where $\tilde{\rho}=\Theta\rho\Theta$. The
eigenvalues of $R:=\rho\tilde{\rho}$ are positive since $R$ is similar to a
Hermitian positive-definite matrix $R=\rho^{1/2}R^\prime\rho^{-1/2}$,
\begin{eqnarray}
R^\prime=\rho^{1/2}\Theta\rho\Theta\rho^{1/2}=
\rho^{1/2} T_1\rho^\ast T_1^\ast\rho^{1/2}=(\rho^{1/2} T_1\rho^{\ast 1/2})
(\rho^{\ast 1/2} T_1^\ast\rho^{1/2})
=(\rho^{1/2} T_1\rho^{\ast 1/2})
(\rho^{1/2} T_1\rho^{\ast 1/2})^\dagger,
\end{eqnarray}
since $\rho=\rho^\dagger$ and $T_1^T=T_1$ i.e.\ $T_1^\dag=T_1^\ast$.

The proof of (\ref{mainformula}) was given by Uhlmann \cite{uhlmann00} (see
also \cite{mckb05}) who also showed that the optimal decomposition may be
constructed out of $2^{n+1}$ vectors, where $2^n<N\le 2^{n+1}$. This is a
generalization of the Wootters construction for the concurrence of two qubit
states \cite{wootters98}.

In the following we will call a (pure or mixed) state `classical' if an
appropriate generalized concurrence (\ref{mixedC}) vanishes.

From the arguments in the last two sections it should be clear that although
vanishing of an appropriate $C_A(\rho)$ can characterize `classical' (eg.\
separable) states in an arbitrary finite dimension, it is only rarely that
its calculation can be reduced to (\ref{mainformula}) \emph{via} some antilinear
$\Theta$. Such situation happens only in low-dimensional cases and this is
the reason why e.g.\ Wootters' construction \cite{wootters98} gives explicit
results only in the two-qubit case. In the next section we will give more
examples supporting this observation.

\section{Examples: separable and uncorrelated states of composite systems}
\label{sec:examples}

\subsection{Entanglement in two-partite systems of distinguishable
particles}

The Hilbert space $\mathcal{H}$ is the tensor product of the Hilbert spaces
of the subsystems, $\mathcal{H}=\mathcal{H}_1\otimes\mathcal{H}_2$ of
dimensions $N_1$ and $N_2$. In this case $A$ is proportional to the
projection on $\mathcal{H}_{12}:=(\mathcal{H}_1\wedge\mathcal{H}_1)
\otimes(\mathcal{H}_2\wedge\mathcal{H}_2)$ \cite{mckb05}. This is a direct
consequence of the fact that for a pure state $\ket{\psi}$ the quantity
\begin{equation}\label{prec}
C^2(\psi)=\bk{\psi}{\psi}^2-\tr\rho_r^2,
\end{equation}
where $\rho_r$ is the reduced (to one of the subsystems) density matrix,
vanishes only for separable $\ket{\psi}$, since it is only in this case that the reduced
density matrix is pure and the trace of its square equals one. It is now a
matter of straightforward calculations to establish that (\ref{prec}) is
proportional to $\bra{\psi\otimes\psi}A\ket{\psi\otimes\psi}$ where $A$ is
the projection on $\mathcal{H}_{12}$.

Let us briefly rederive this result using Schmidt decomposition, since later
we would like to stress analogies with other examples of `classical' systems.
An arbitrary two-partite pure state can be written as
\begin{equation}\label{2partpure}
\ket{\psi}=\sum_{i,j}c_{ij}\, \ket{e_i}\otimes\ket{f_j},
\end{equation}
where $\{\ket{e_i}\}|_{i=1}^{N_1}$ and $\{\ket{f_j}\}_{j=1}^{N_2}$ are some
orthonormal bases in $\mathcal{H}_1$ and $\mathcal{H}_2$ and $c$ some complex
matrix. An arbitrary complex matrix $c$ can be transformed to a diagonal one
with non-negative entries by multiplication from left and right by unitary
matrices, $c\mapsto UcV$, \cite{horn85},  which amounts to local unitary
changes of bases, $\ket{e_i}=\sum_k U_{ki}\ket{e_k^\prime}$,
$\ket{f_j}=\sum_l V_{jl}\ket{f_l^\prime}$. Hence, upon an appropriate choice
of $U$ and $V$ we obtain $\psi$ in its Schmidt form
\begin{equation}\label{schmidt}
\ket{\psi}=\sum_k l_k\ket{e_k^\prime}\otimes\ket{f_k^\prime}, \quad l_k>0,
\quad \sum_k l_k^2=1,
\end{equation}
where the last condition stems from the normalization $\tr\kb{\psi}{\psi}=1$.
The state $\ket{\psi}$ is a simple tensor (i.e.\ is separable) if and only if
only one among $l_k$ does not vanish. It is now easy to show that this
condition is equivalent to the vanishing of $C$ given by (\ref{prec}).

The classical pure states are thus the separable ones. Hence the classical
mixed states are those which are decomposable into pure separable states,
i.e.\ (mixed) separable states.

For two qubits, $N_1=2=N_2$, the matrix $A$ reads
 \begin{equation}\label{Aqubits}
A=
\left[ \begin {array}{rrrrrrrrrrrrrrrr}
0&0&0&0&0&0&0&0&0&0&0&0&0&0&0&0\\\noalign{\medskip}0&0&0&0&0&0&0&0&0&0&0&0&0&0&0&0
\\\noalign{\medskip}0&0&0&0&0&0&0&0&0&0&0&0&0&0&0&0
\\\noalign{\medskip}0&0&0&1&0&0&-1&0&0&-1&0&0&1&0&0&0
\\\noalign{\medskip}0&0&0&0&0&0&0&0&0&0&0&0&0&0&0&0
\\\noalign{\medskip}0&0&0&0&0&0&0&0&0&0&0&0&0&0&0&0
\\\noalign{\medskip}0&0&0&-1&0&0&1&0&0&1&0&0&-1&0&0&0
\\\noalign{\medskip}0&0&0&0&0&0&0&0&0&0&0&0&0&0&0&0
\\\noalign{\medskip}0&0&0&0&0&0&0&0&0&0&0&0&0&0&0&0
\\\noalign{\medskip}0&0&0&-1&0&0&1&0&0&1&0&0&-1&0&0&0
\\\noalign{\medskip}0&0&0&0&0&0&0&0&0&0&0&0&0&0&0&0
\\\noalign{\medskip}0&0&0&0&0&0&0&0&0&0&0&0&0&0&0&0
\\\noalign{\medskip}0&0&0&1&0&0&-1&0&0&-1&0&0&1&0&0&0
\\\noalign{\medskip}0&0&0&0&0&0&0&0&0&0&0&0&0&0&0&0
\\\noalign{\medskip}0&0&0&0&0&0&0&0&0&0&0&0&0&0&0&0
\\\noalign{\medskip}0&0&0&0&0&0&0&0&0&0&0&0&0&0&0&0\end {array}
 \right].
\end{equation}
It has only one nonvanishing eigenvalue (equal to $4$) with the eigenvector
\begin{equation}\label{vqubits}
\ket{w_0}=\left[ \begin {array}{cccccccccccccccc}
0&0&0&-1&0&0&1&0&0&1&0&0&-1&0&0&0
\end {array} \right]^T.
\end{equation}
The corresponding matrix $T$ reads
\begin{equation}\label{Tqubits}
T=
\left[ \begin {array}{rrrr}
0&0&0&-1\\\noalign{\medskip}0&0&1&0
\\\noalign{\medskip}0&1&0&0
\\\noalign{\medskip}-1&0&0&0\end {array}
 \right]=\sigma_y\otimes\sigma_y,
\end{equation}
and the concurrence is given by the Wootters formula,
\begin{equation}\label{wootters}
C(\rho)=\max\Big\{0,\mu_1-\sum_{j=2}^4\mu_j\Big\},
\end{equation}
Here $\mu_j$ are the singular values of
$\tau_{ij}=\bra{\xi_i}T\ket{\xi_j^\ast}$ and $\ket{\xi_i}$, $i=1,\ldots,4$
--- the subnormalized eigenvectors of $\rho$, or equivalently, $\mu_i$ are
the square roots of the eigenvalues of $\rho\tilde{\rho}$, where
$\tilde{\rho}= T\rho^\ast T$. The optimal decomposition can be constructed
out of four vectors, which is also clear from Wootters' original
construction.

\subsection{Entanglement in two-fermion systems}
Correlations in systems of two fermions were investigated in
\cite{sckll01,eckert02}, where all relevant definitions and proofs can be
found. Here we only briefly review the most important findings. In this case
$\mathcal{H}$ is the antisymmetric part of the tensor product of two copies
of the single-particle Hilbert space $\mathcal{H}_{2K}$ of an even dimension
$2K=2S+1$, where $S$ is the spin of each particle,
$\mathcal{H}=\mathcal{A}(\mathcal{H}_{2K}\otimes\mathcal{H}_{2K})$, hence
$\mathcal{H}$ has dimension $N=2K(2K-1)/2=S(2S+1)$. An arbitrary pure state
can be represented in the form
\begin{equation}\label{fermions}
\ket{\psi}=\sum_{i,j=1}^{2S+1}w_{ij}f_i^\dagger f_j^\dagger\ket{0},
\end{equation}
where $f_i^\dagger$ are fermionic creation operators and $\ket{0}$ --- the
vacuum state and $w$ --- a complex antisymmetric matrix. A unitary
transformation $U$ of the single particle space $\mathcal{H}_S$ leads to the
transformation $f_i^\dagger\mapsto\sum_{ji}U_{ji}f_j^\dagger$ and,
consequently,
\begin{equation}\label{mapw}
w\mapsto UwU^T.
\end{equation}
The analogue of the Schmidt decomposition is now provided by the theorem
stating that an arbitrary, complex matrix $w$ can be brought by an appropriate
transformation (\ref{mapw}) to a block-diagonal, canonical form,
\begin{equation}\label{wcan}
w^\prime=\mathrm{diag}\left[Z_1,\ldots,Z_r,Z_0\right], \quad
Z_i=\left[
\begin{array}{cc}
  0   & z_i \\
 -z_i &  0  \\
\end{array}
\right],
\end{equation}
with $z_i\ne 0$ and $Z_0$ - a null matrix \cite{horn85}. The number $r$ of
non-vanishing $2\times 2$ blocks is called the \emph{Slater rank} of
$\ket{\psi}$ \cite{sckll01}.

Pure states with the minimal (i.e.\ equal to one) Slater rank exhibit the
minimal allowable quantum correlations \cite{sckll01}, so they are the
candidates for the `classical' states. Consequently, mixed states are
`classical' when they can be decomposed into a convex combination of pure
states with Slater rank one. The maximal Slater rank of the components of a
pure state decomposition of a mixed state $\rho$ minimized over all possible
decomposition is called the \emph{Slater number}, by analogy to the Schmidt
number defined in a similar manner for distinguishable particles. Using this
notion we identify `classical' mixed states with those of the Slater number
one.

In order to characterize the `classical' states we make use of a lemma proved
in \cite{eckert02}. It provides a general criterion for a pure state in the
form (\ref{fermions}) to have a Slater rank not exceeding a prescribed
value. In particular it states that a two fermion state in
$\mathcal{H}=\mathcal{A}(\mathcal{H}_{2K}\otimes\mathcal{H}_{2K})$ has the
Slater rank one if and only if for all
$1\le\alpha_1<\cdots<\alpha_{2(K-2)}\le 2K$
\begin{equation}\label{lemmafermions}
\sum_{i,j,k,l=1}^{2K}w_{ij}w_{kl}\varepsilon^{ijkl\alpha_1\ldots\alpha_{2(K-2)}}
=0
\end{equation}
 where $\varepsilon^{i_1\ldots i_{2K}}$ is the totally antisymmetric unit
tensor in $\mathcal{H}_{2K}$, which is obviously equivalent to
\begin{equation}\label{bil-ferm1}
\sum_{1\le\alpha_1<\cdots<\alpha_{2(K-2)\le 2K}}\left
|\sum_{i,j,k,l=1}^{2K}w_{ij}w_{kl}\varepsilon^{ijkl\alpha_1\ldots
\alpha_{2(K-2)}}\right|^{\, 2}=0.
\end{equation}
Each of the terms under the absolute value sign is the Pfaffian
\footnote{Pfaffian of an antisymmetric even-dimensional matrix $w$ is the
polynomial whose square is the determinant of $A$ \cite{serre02}.} of the
$4\times 4$ matrix obtained from $w$ by deleting the rows an columns with
numbers $\alpha_1,\alpha_2,\ldots,\alpha_{2(K-2)}$. It is a matter of simple
calculations that the Pfaffian $\mathrm{Pf}(X)$ of a $4\times 4$ matrix $X$
reads
\begin{equation}\label{pfaffian}
\mathrm{Pf}(X)=X_{12}X_{34}-X_{13}X_{24}+X_{14}X_{23},
\end{equation}
and
\begin{equation}\label{pfaffian1}
\left|\mathrm{Pf}(X)\right|^2=\bra{\mathbf{x}\otimes\mathbf{x}}A_{Pf}
\ket{\mathbf{x}\otimes\mathbf{x}},
\end{equation}
where $\ket{\mathbf{x}}$ is the six-dimensional vector
\begin{equation}\label{x}
\ket{\mathbf{x}}=\left[
\begin{array}{c}
 X_{12} \\
 X_{13} \\
 X_{14} \\
 X_{23} \\
 X_{24} \\
 X_{34} \\
\end{array}
\right],
\end{equation}
and, in the standard basis $\ket{e_1},\ldots,\ket{e_6}$ in $\mathbb{C}^6$,
\begin{equation}\label{APf}
A_{Pf}=\sum_{i,j=1}^6\kb{e_i}{e_j}\otimes T\kb{e_i}{e_j} T^\dagger,
\end{equation}
with
\begin{equation}\label{T-fermion}
T=
\left[ \begin {array}{cccccc} 0&0&0&0&0&1\\\noalign{\medskip}0&0&0&0&-1&0\\\noalign{\medskip}0&0&0&1&0&0\\\noalign{\medskip}0&0&1&0&0&0
\\\noalign{\medskip}0&-1&0&0&0&0\\\noalign{\medskip}1&0&0&0&0&0
\end {array} \right].
\end{equation}

Substituting for each term in the sum in (\ref{bil-ferm1}) an appropriate
expression (\ref{pfaffian1}), involving in each summand a different set of
entries of $w$ with the corresponding matrix $A_{Pf}$, we rewrite the
condition (\ref{bil-ferm1})
\begin{equation}\label{gen-ferm}
C_A^2(\psi):=\bra{\mathbf{w}\otimes\mathbf{w}}A\ket{\mathbf{w}\otimes\mathbf{w}}=0.
\end{equation}
Here $A$ is a $S^2(2S+1)^2\times S^2(2S+1)^2$ matrix acting in
$\mathcal{H}\otimes\mathcal{H}$, built up from the submatrices $A_{Pf}$ of
each summand, and $\ket{\mathbf{w}}$ is the $2S+1$ dimensional vector of
independent entries of the matrix $w$ representing in this way the state
vector $\ket{\psi}$ (\ref{fermions}). The condition (\ref{gen-ferm})
constitutes the appropriate non-entanglement criterion for pure fermionic
states in the form of a bilinear generalized concurrence (\ref{CA}).

For the lowest-dimensional, nontrivial case, i.e.\ $S=3/2$, $A$ is a
$36\times 36$ matrix (skipped to save the place) with only one non-vanishing
eigenvalue. The corresponding eigenvector has only six non-vanishing elements
and the corresponding matrix $T$ (\ref{Ta}) is given by (\ref{T-fermion}), as
it is also clear from (\ref{Ajam}) and (\ref{APf}). The result is equivalent
to the one of \cite{eckert02} \textit{modulo} a change of basis in
$\mathbb{C}^6$: $\ket{e_1^\prime}=\ket{e_1}$, $\ket{e_2^\prime}=\ket{e_2}$,
$\ket{e_3^\prime}=(\ket{e_3}+\ket{e_4})/\sqrt{2}$,
$\ket{e_4^\prime}=\ket{e_5}$, $\ket{e_5^\prime}=\ket{e_6}$, and
$\ket{e_6^\prime}=(\ket{e_3}-\ket{e_4})/\sqrt{2}$.

The generalized concurrence $C_A(\rho)$ for mixed states is the \emph{Slater
correlation measure} \cite{sckll01} and is given by
\begin{equation}
 C_\mathrm{Sl}(\rho)=\max\Big\{0,\mu_1-\sum_{j=2}^6\mu_j\Big\}.
\end{equation}
The definition of $\mu_i$ is analogous to the one used in the preceding
example, the only changes are in the dimensionality and the definition of
$T$. As previously $\mu_i$ are equal to the square roots of the eigenvalues
of $\rho\tilde{\rho}$, with $\tilde{\rho}=T\rho^\ast T$, which coincides with
the results of \cite{sckll01,eckert02}. From the result of Uhlmann we infer
that the optimal decomposition can be achieved with $8$ vectors.


\subsection{Entanglement in two-boson systems}
Quantum correlations in bosonic systems were thoroughly investigated in
\cite{eckert02}. The line of thought follows \emph{mutatis mutandis}
considerations in the fermionic case. For two particles the Hilbert space is
the symmetric part of the tensor product of the single-particle space,
$\mathcal{H}=\mathcal{S}(\mathcal{H}_M \otimes\mathcal{H}_M)$,
$\dim\mathcal{H}_M=M$, $\dim\mathcal{H}=M(M+1)/2$. A pure state can be
written in the form
\begin{equation}\label{bosons}
\ket{\psi}=\sum_{i,j=1}^M b_i^\dagger b_j^\dagger v_{ij}\ket{0},
\end{equation}
with bosonic creation operators $b_i^\dagger$ and a symmetric complex matrix
$v$ transforming upon unitary map $U$ in the single particle state according
to $b_i^\dagger\mapsto \sum_i U_{ij}b_j^\dagger$ and, consequently,
\begin{equation}\label{mapboson}
v\mapsto UvU^T.
\end{equation}
As previously, an appropriate theorem from the linear algebra \cite{horn85}
provides a possibility of using (\ref{mapboson}) to transform $v$ to its
diagonal, canonical form
\begin{equation}\label{canboson}
v^\prime=\mathrm{diag}(z_1,\ldots,z_r,0,\ldots,0), \quad z_i\ne 0.
\end{equation}
The number $r$ of nonvanishing $z_i$ is dubbed \emph{bosonic Slater rank},
and states with the minimal $r=1$ exhibit the minimal possible amount of
purely quantum correlations \cite{eckert02}. Accordingly these are our
`classical' pure states, whereas `classical' mixed states are those which can
be decomposed into a convex combination of pure states with the Slater rank
equal to one. As in the previously considered cases one can define the \emph{
bosonic Slater number} of a mixed state as a minimum over all convex
decompositions into pure states of the maximal Slater rank among the members
of a decomposition. In this terminology, a mixed is `classical' if and only
if its bosonic Slater number equals one.

As in the case of fermions there exists a bilinear characterization of pure
states with the bosonic Slater rank equal to one \cite{eckert02}. For them
\begin{equation}\label{lemmaboson}
\sum_{ijkl=1}^M v_{ij}v_{kl}\varepsilon^{ik\alpha_1\ldots\alpha_{M-2}}
\varepsilon^{jl\alpha_1\ldots\alpha_{M-2}}=0
\end{equation}
for all $1\le\alpha_1<\cdots<\alpha_{M-2}\le M$, which can be transformed to
the desired form (\ref{CA}) along, essentially, the same lines as in the
fermionic case. To this end we rewrite (\ref{lemmaboson}) in the form
\begin{equation}\label{bil-bos1}
\sum_{1\le\alpha_1<\cdots<\alpha_{M-2}\le M}\left|
\sum_{i,j,k,l=1}^M v_{ij}v_{kl}\varepsilon^{ik\alpha_1\ldots\alpha_{M-2}}
\varepsilon^{jl\alpha_1\ldots\alpha_{M-2}} \right|^{\,2}=0.
\end{equation}
To cut short the connection to the conventions used in \cite{eckert02} it is
convenient to represent the vector $\ket{\psi}$ (\ref{bosons}) in
$\mathcal{H}=\mathbb{C}^{M(M+1)/2}$ by the vector
\begin{equation}\label{v-boson}
\ket{\mathbf{v}}=\left[
\begin{array}{c}
 v_{11}^\prime \\
    v_{12}     \\
    \vdots     \\
 v_{MM}^\prime \\
\end{array}
\right]
\end{equation}
of the independent entries of the symmetric matrix $v$, i.e.\ $v_{ij}$, $i\le
j$, scaling, however, the diagonal entries by $1/\sqrt{2}$,
$v_{jj}^\prime=v_{jj}/\sqrt{2}$. The way of representing $\ket{\psi}$ in
terms of $v$ is, obviously, a matter of convenience, as long as linearity of
the representation is observed. The chosen one, besides being in accordance
with \cite{eckert02}, has the advantage of giving the same `weight' to
diagonal and off-diagonal entries of $v$.

Each summand under the absolute value sign in (\ref{bil-bos1}) involves only
a $2\times 2$ submatrix of $v$,
\begin{equation}\label{bos-subm}
v(a,b)=\left[
\begin{array}{cc}
 v_{aa} & v_{ab} \\
 v_{ab} & v_{bb}  \\
\end{array}
\right], \quad 1\le a,b\le M,
\end{equation}
and equals, up to an irrelevant sign, $2v_{aa}v_{bb}-v_{ab}^2=v_{aa}^\prime
v_{bb}^\prime-v_{ab}^2$. This, in turn, can be rewritten in the form
$\bra{\mathbf{\tilde{v}}\otimes\mathbf{\tilde{v}}}A_{bos}
\ket{\mathbf{\tilde{v}}\otimes\mathbf{\tilde{v}}}$, where

\begin{equation}\label{T-bos}
\ket{\mathbf{\tilde{v}}}=\ket{\mathbf{\tilde{v}(a,b)}}=\left[
\begin{array}{c}
 v_{aa}^\prime \\
    v_{ab}     \\
 v_{bb}^\prime \\
\end{array}
\right], \quad
A_{bos}=\sum_{i,j=1}^3\kb{e_i}{e_j}\otimes T\kb{e_i}{e_j} T^\dagger, \quad
T=\left[ \begin {array}{ccc}
0&0&1\\\noalign{\medskip}0&-1&0\\\noalign{\medskip}1&0&0\end {array}
\right],
\end{equation}
and $\left\{\ket{e_1},\ket{e_2},\ket{e_3}\right\}$ is the standard basis in
$\mathbb{C}^3$. Again, the non-entanglement condition for pure states
(\ref{bil-bos1}) can be rewritten as
\begin{equation}\label{gen-bos}
C_A^2(\psi):=\bra{\mathbf{v}\otimes\mathbf{v}}A\ket{\mathbf{v}
\otimes\mathbf{v}}=0,
\end{equation}
by combining the appropriate $6\times 6$ matrices $A_{bos}$ into a
$\left(M(M+1)/2\right)^2\times \left(M(M+1)/2\right)^2$ matrix acting in
$\mathcal{H}\otimes\mathcal{H}$

In the lowest-dimensional, nontrivial case $M=2$, $N=4$ we find \footnote{The
somehow awkward fact that the dimension of the single particle space is even
should not be misleading, it is not determined by the total spin, but rather
by the number of available single-particle states.}
\begin{equation}\label{bosonicA}
A=\left[ \begin {array}{rrrrrrrrr}
0&0&0&0&0&0&0&0&0\\\noalign{\medskip}0&0&0&0&0&0&0&0&0\\\noalign{\medskip}0&0&1&0&-1&0&
1&0&0\\\noalign{\medskip}0&0&0&0&0&0&0&0&0\\\noalign{\medskip}0&0&-1&0
&1&0&-1&0&0\\\noalign{\medskip}0&0&0&0&0&0&0&0&0\\\noalign{\medskip}0&0
&1&0&-1&0&1&0&0\\\noalign{\medskip}0&0&0&0&0&0&0&0&0
\\\noalign{\medskip}0&0&0&0&0&0&0&0&0\end {array} \right].
\end{equation}
The matrix $A$ has only one non-vanishing eigenvalue with the eigenvector
\begin{equation}
\ket{w_0}=\left[ \begin {array}{ccccccccc} 0&0&1&0&-1&0&1&0&0\end {array}
\right]^T,
\end{equation}
to which there corresponds \emph{via} (\ref{Ta}) the matrix $T$ given in
(\ref{T-bos}). In a perfect analogy with the previous examples the
generalized concurrence $C_A(\rho)$ is given in terms of the square roots of
the eigenvalues of $\rho T\rho^\ast T$,
\begin{equation}
C_A(\rho)=\max\Big\{0,\mu_1-\sum_{j=2}^3\mu_j\Big\},
\end{equation}
and coincides with the \emph{bosonic correlation measure} \cite{eckert02}.
The optimal decomposition involves at most $4$ vectors.

\section{Spin coherent states}

Classical spin states were defined in \cite{giraud08} as those which can be
decomposed into a probabilistic mixture of pure spin coherent states. In
order to mimic the previous construction we will use the following
characterization of pure coherent states, following directly from their
minimal-uncertainty property \cite{klyachko08}.
\begin{theo}
A spin-$S$ state $\ket{\psi}\in\mathcal{H}$ is coherent if and only if
\begin{equation}\label{klyachko}
(L_1\otimes L_1+L_2\otimes L_2+L_3\otimes L_3)\ket{\psi\otimes\psi}
=S^2\ket{\psi\otimes\psi},
\end{equation}
where $L_i$ are operators of the spin components in the spin $S$
representation of the rotation group.
\end{theo}

This is obviously equivalent to
\begin{equation}
\bra{\psi\otimes\psi}A\ket{\psi\otimes\psi}=0,\quad
A=I-L_1\otimes L_1-L_2\otimes L_2-L_3\otimes L_3.
\end{equation}
For the lowest-dimensional nontrivial case $S=1$ we choose
\begin{equation}\label{Ls}
L_1=\left[ \begin {array}{rrr}
0&0&0\\\noalign{\medskip}0&0&-i\\\noalign{\medskip}0&i&0\end {array}
\right],
\quad
L_2=\left[ \begin {array}{rrr}
0&0&i\\\noalign{\medskip}0&0&0\\\noalign{\medskip}-i&0&0\end {array}
\right],
\quad
L_3=\left[ \begin {array}{rrr}
0&-i&0\\\noalign{\medskip}i&0&0\\\noalign{\medskip}0&0&0\end {array}
\right].
\end{equation}
Then
\begin{equation}\label{coherentA}
A=
\left[ \begin {array}{rrrrrrrrr}
1&0&0&0&1&0&0&0&1\\\noalign{\medskip}0&1&0&-1&0&0&0&0&0\\\noalign{\medskip}0&0&1&0&0&0&
-1&0&0\\\noalign{\medskip}0&-1&0&1&0&0&0&0&0\\\noalign{\medskip}1&0&0&0
&1&0&0&0&1\\\noalign{\medskip}0&0&0&0&0&1&0&-1&0\\\noalign{\medskip}0&0
&-1&0&0&0&1&0&0\\\noalign{\medskip}0&0&0&0&0&-1&0&1&0
\\\noalign{\medskip}1&0&0&0&1&0&0&0&1\end {array} \right].
\end{equation}
The matrix $A$ has eigenvalues $3,2,2,2,0,0,0,0,0$, hence, at first sight,
the previous construction fails. But by inspection of the eigenvectors we
find that the corresponding matrices $T_\alpha$ read
\begin{equation}\label{T1234}
T_1=I, \quad T_2=iL_1, \quad T_3=iL_2, \quad T_4=iL_3,
\end{equation}
hence, due to the antisymmetry of $T_i$, $i=2,3,4$, we have
$\bra{\psi}T_i\ket{\psi^\ast}=0$ for $i=2,3,4$, and an arbitrary $3$-dim
vector $\ket{\psi}$. Thus (cf. (\ref{psiApsigen}))
\begin{equation}
C_A(\psi)=|\bk{\psi}{\psi^\ast}|,
\end{equation}
and
\begin{equation}\label{cohCA}
C_A(\rho)=\max\Big\{0,\mu_1-\sum_{j=2}^3\mu_j\Big\},
\end{equation}
where $\mu_i$ are the singular values of $\tau_{ij}=\bk{\xi_i}{\xi_j^\ast}$
and $\ket{\xi_i}$ are the subnormalized eigenvectors of $\rho$ (or,
equivalently, $\mu_i$ are the roots of the eigenvalues of $R=\rho\rho^\ast$).
The optimal decomposition can be achieved with at most \textbf{four} vectors.

The criterion of classicality based on (\ref{cohCA}) can be formulated
directly in terms of orthogonal invariants (traces of powers) of $R$. To this
end consider a $3\times 3$ density matrix i.e.\ $\rho=\rho^\dagger$,
$\tr\rho=1$ and $\rho$ --- non-negatively definite. In the following we
assume that $\rho$ is nondegenerate, $\det\rho>0$, we also will suplement
this by some other (see below) non-degeneracy conditions (all degenerate
cases, in general simpler in treatment, can be considered along the same
lines).

We decompose $\rho$ into the real and imaginary part, $\rho=\rho_R+i\rho_I$.
The hermiticity of $\rho$ implies $\rho_R^t=\rho_R$, $\rho_I^T=-\rho_I$. The
real part $\rho_R$ can be thus diagonalized by an orthogonal transformation
which leaves the antisymmetry of $\rho_I$ unaltered. After that $\rho$ takes
the form
\begin{equation}\label{rhocan}
 \rho=\left[ \begin {array}{ccc} {\lambda_1}&-i{ v_3}&i{
v_2}\\\noalign{\medskip}i{ v_3}&{\lambda_2}&-i{ v_1}
\\\noalign{\medskip}-i{ v_2}&i{ v_1}&{\lambda_3}\end {array}
 \right].
\end{equation}
Since $\rho$ is non-negatively definite we have $\lambda_i\ge 0$. In the
following we assume a further non-degeneracy condition $\lambda_i>0$. We have
also $\lambda_1+\lambda_2+\lambda_3=\tr\rho=1$.

The matrix $R$ reads now:
\begin{equation}
R=\rho\rho^\ast=
\left[ \begin {array}{ccc}
{{\lambda_1}}^{2}-{{ v_2}}^{2}-{{v_3}}^{2}&
v_1v_2+iv_3({\lambda_1}-{\lambda_2})&
v_1v_3+iv_2({\lambda_3}-{\lambda_1})
\\\noalign{\medskip}
v_1v_3+iv_2({\lambda_3}-{\lambda_1})&
{{\lambda_2}}^{2}-{{ v_1}}^{2}-{{ v_3}}^{2}&
v_2v_3+iv_1({\lambda_2}-{\lambda_3})
\\\noalign{\medskip}
v_1v_3+iv_2({\lambda_3}-{\lambda_1})&
v_2v_3+iv_1({\lambda_2}-{\lambda_3})
&{{\lambda_3}}^{2}-{{ v_2}}^{2}-{{ v_1}}^{2}
\end {array} \right].
\end{equation}
From the previous arguments we know that $R$ has a real non-negative
spectrum. We denote the eigenvalues of $R$ by $\mu_1^2,\mu_2^2$, and
$\mu_3^2$, and assume $\mu_1>\mu_2>\mu_3>0$ (again leaving apart possible
degeneracies) and define
\begin{equation}\label{x1}
x_1=\mu_1-\mu_2-\mu_3, \quad x_2=\mu_2-\mu_1-\mu_3, \quad
x_2=\mu_3-\mu_1-\mu_2, \quad x_4=\mu_1+\mu_2+\mu_3.
\end{equation}
Hence $x_2=-(\mu_1-\mu_2)-\mu_3<0$, $x_3=-(\mu_1-\mu_3)-\mu_2<0$, and
$x_4>0$. It follows thus that the polynomial
\begin{equation}\label{P}
P(x)=(x-x_1)(x-x_2)(x-x_3)(x-x_4)
\end{equation}
has at least two negative roots $x_2$ and $x_3$ and at least one positive
root $x_4$. The sign of the remaining one $x_1$ depends on $\lambda_i$ and
$v_i$.

Straightforward calculations give
\begin{equation}\label{P1}
P(x)=x^4-(4\tr R)x^2-8(\det R)^{1/2}x+2\,\tr(R^2)-(\tr R)^2
\end{equation}
If $2\,\tr(R^2)-(\tr R)^2<0$ then the consecutive signs of the coefficients
of $P(x)$ read $+---$, hence from the Descartes rule of signs it has at most
one positive root and consequently $x_1=\mu_1-\mu_2-\mu_3<0$. On the other
hand for $2\,\tr(R^2)-(\tr R)^2>0$ the signs of the coefficients of $P(-x)$
are $+-++$, hence $P(-x)$ has at most two positive roots, i.e.\ $P(x)$ has at
most two negative roots and, consequently, $x_1$ is positive. (The validity
of this assertion can be easily established by considering the graph of
$P(x)$ under the conditions that it crosses the $x$-axis in at least two
negative and at least one positive point). Summarizing: the sign of
$\mu_1-\mu_2-\mu_3$ coincides with the sign of $2\,\tr(R^2)-(\tr R)^2$, or in
other words, $\rho$ is `classical' if and only if $2\,\tr(R^2)-(\tr R)^2\le
0$. (The equality in the last formula appears when we relax the
non-degeneracy conditions).

It can be proved that (\ref{cohCA}) is equivalent to the criterion of
\cite{giraud08} for the classicality of spin-1 coherent states. The latter is
formulated as follows. We rewrite $\rho$ in the form
\begin{equation}\label{W}
\rho=\frac{1}{2}\left(I-W+\mathbf{u}\cdot\mathbf{L}\right),
\end{equation}
where $\mathbf{u}$ is a real vector and $\mathbf{L}=(L_1,L_2,L_3)$  with
$L_i$ given by (\ref{Ls}). The matrix $Z$ of Giraud et.al. is then defined as
$Z_{ij}=W_{ij}-u_iu_j$. The state is classical iff $Z$ is positive definite.
The proof of the equivalence of both criteria is given in the Appendix.

\section{`Classical' states as generalized coherent states}

As already stated in the introduction similarities of all above outlined
constructions have their common origin in the fact that `classical' states
are special orbits of the Lie group of symmetries characteristic for a
particular problem. To make this statement more precise let us recall a few
facts from the theory of group representations needed for the definition of
generalized coherent states \cite{hall:03} . Let $K$ be a compact semisimple
Lie group, $\mathfrak{k}$ its Lie algebra and
$\mathfrak{g}=\mathfrak{k}+i\mathfrak{k}$ - the complexification of
$\mathfrak{k}$. As a convenient example we can take $K=SU(N)$,
$\mathfrak{k}=\mathfrak{su}_N$, and
$\mathfrak{g}=\mathfrak{sl}_N(\mathbb{C})$. In addition we can imagine all of
them as sets of complex matrices in the defining representation in the
complex space $\mathbb{C}^N$, then $K=SU(N)$ is the set of unitary $N\times
N$ matrices with determinant 1, $\mathfrak{k}=\mathfrak{su}_N$ --- the set of
traceless antihermitian $N\times N$ matrices, and
$\mathfrak{sl}_N(\mathbb{C})$ --- the set of complex, traceless $N\times N$
matrices.

The Lie algebra $\mathfrak{g}$ can be decomposed into the direct sum
\begin{equation}
\mathfrak{g}=\mathfrak{n}_-\oplus\mathfrak{t}\oplus\mathfrak{n}_+,
\end{equation}
where $\mathfrak{t}$ is the Cartan subalgebra of $\mathfrak{g}$ (its maximal
commutative subalgebra dimension of which is called the rank of
$\mathfrak{g}$), whereas $\mathfrak{n}_\pm$ are particular nilpotent
subalgebras of $\mathfrak{g}$. In our $SU(N)$ example realized in
$\mathbb{C}^N$ we can choose $\mathfrak{t}$ as the set of traceless diagonal
matrices and $\mathfrak{n}_+$ and $\mathfrak{n}_-$ as, respectively, upper-
and lower-triangular matrices.

As linear spaces $\mathfrak{n}_+$ and $\mathfrak{n}_-$ are direct sums of the
\emph{root spaces}
\begin{equation}
\mathfrak{n}_+=\bigoplus_\alpha\mathfrak{g}_\alpha,\quad
\mathfrak{n}_-=\bigoplus_\alpha\mathfrak{g}_{-\alpha}.
\end{equation}
Each $\mathfrak{g}_\alpha$ is one-dimensional and is uniquely determined by
the commutation relations of its elements with the elements in $\mathfrak{t}$
\begin{equation}\label{root}
[H,X]=\alpha(H)X, \quad H\in\mathfrak{t},\quad X\in\mathfrak{g}_\alpha,
\end{equation}
where $\alpha(\cdot)$ is an appropriate linear form on $\mathfrak{t}$. If we
choose some basis $\{H_i\}$, $(i=1,\ldots,r=\dim\mathfrak{t})$ in
$\mathfrak{t}$, we can calculate (\ref{root}) for $H=H_i$ obtaining vectors
(roots) $\boldsymbol{\alpha}$ with the components
$\boldsymbol{\alpha}_i=\alpha(H_i)$. They span the Euclidean space of
dimension equal to the rank of $\mathfrak{g}$, in which we can choose a basis
consisting of \emph{positive simple roots} --- each root
$\boldsymbol{\alpha}$ is a linear combination of them with only non-negative
(positive roots) or non-positive (negative roots) coefficients. The algebra
$\mathfrak{g}$ is uniquely determined by its positive roots (or,
equivalently, by its positive simple roots). The \emph{positive (negative)
root vectors} are, by definition, the elements $X_{\pm\alpha}$, of
$\mathfrak{g}_{\pm\alpha}$, fulfilling, according to (\ref{root}),
$[H_i,X_{\pm\alpha}]=\pm\alpha_iX_{\pm\alpha}$. There is one-to-one
correspondence between the negative and positive root vectors.

On a semisimple Lie algebra $\mathfrak{g}$ there exists a nondegenerate
bilinear form (the Killing form) defined as
\begin{equation}\label{killing}
(X,Y)=\tr(\mathrm{ad}X\cdot\mathrm{ad}Y),
\end{equation}
where $\mathrm{ad}X$ is the linear transformation of $\mathfrak{g}$ into
itself given by $\mathrm{ad}X(Y)=[X,Y]$. We can use the Killing form to fix the
normalization of root vectors and the elements $H_i$ by choosing
$(X_\alpha,X_{-\alpha})=1$ and $(H_i,H_j)=\delta_{ij}$.

From the root vectors and the elements $H_i$ we can construct the second
order \emph{Casimir operator}
\begin{equation}\label{casimir}
C_2:=\sum_{\alpha>0}(X_\alpha X_{-\alpha}+X_{-\alpha} X_{\alpha})
+\sum_{i=1}^rH_i^2,
\end{equation}
which, as straightforward calculations show, commutes with every element of
$\mathfrak{g}$. With the help of $C_2$ one can conveniently characterize the
set of coherent states.

The group $K$ as well as the algebras $\mathfrak{k}$ and $\mathfrak{g}$ can
be irreducibly represented not only in the defining space $\mathbb{C}^N$, as
we did with $SU(N)$, but also in spaces of other dimensions. If we choose as
the representation space $\mathcal{H}=\mathbb{C}^M$, we can again think of
representatives of $K$, $\mathfrak{g}$, etc. as sets of matrices, so we will
use the same letter $\pi$ to denote the homomorphism between the sets of
matrices in $\mathbb{C}^N$ and $\mathbb{C}^M$ defining the considered
irreducible representation. Thus we denote by $\pi(H)$ the matrix
representing in $\mathbb{C}^M$ the  element $H$ of, say, $\mathfrak{g}$, or
by $\pi(U)$ a representative of $U\in K$ \textit{etc}.

For each irreducible representation of a semisimple $\mathfrak{g}$ there
exists a vector $\ket{v_{max}}$ in $\mathcal{H}$ which is a common
eigenvector of all $\pi(H_i)$ annihilated by $\pi(X)$ for
$X\in\mathfrak{n}_+$, i.e.\
\begin{equation}\label{hiweight}
\pi(H_i)\ket{v_{max}}=\lambda_i \ket{v_{max}}, \quad \pi(X)\ket{v_{max}}=0,
\quad \mathrm{for\  }
X\in\mathfrak{n}_+.
\end{equation}
The vector $\ket{v_{max}}$ is called the highest weight vector of the
irreducible representation. An irreducible representation of $\mathfrak{g}$,
as well as $K$, is uniquely determined by the vector
$\boldsymbol\lambda=(\lambda_1,...,\lambda_r)$.

Generalized coherent states for the group $K$, or more precisely, for its
irreducible representation in $\mathcal{H}$, are obtained by applying to the
highest weight vector $\ket{v_{max}}$ all possible representatives $\pi(U)$,
$U\in K$ \footnote{Perelomov \cite{perelomov86} defines generalized coherent
states by choosing arbitrary fixed vector $\ket{u}$ in $\mathcal{H}$ in place
of $v$. The states obtained by action on $\ket{v_{max}}$ are then called
`closest to classical'.}. In all cases relevant for the present
considerations, we can assume that the representation in question is unitary,
so the action of the group \textit{via} its representation does not change
the lengths of vectors in $\mathcal H$, and we can assume $\ket{v_{max}}$ to
be normalized to the unit length. Since in quantum mechanics we identify
states differing by a phase factor, it is more convenient to look at the
coherent states as an orbit of $K$ in the projective space
$\mathbb{P}\mathcal{H}$, ie. the space of the equivalence classes of vectors
differing by a multiplicative constant. Let us denote the equivalence class
of some $\ket{w}\in\mathcal{H}$ by $[\ket{w}]$
\begin{equation}\label{prquiv}
[\ket{w}]=\{\ket{w^\prime}\in\mathcal{H}: \ket{w^\prime}=c\ket{w},\ c\in\mathbb{C}\}.
\end{equation}
Since the representation $\pi$ acts in a natural way in
$\mathbb{P}\mathcal{H}$,
\begin{equation}\label{prepr}
\pi(U)[\ket{w}]=[\pi(U)\ket{w}],
\end{equation}
we may identify the manifold of coherent states $\mathcal{C}_S$ with the
orbit of $K$ in $\mathbb{P}\mathcal{H}$ through the equivalence class (the
complex line) of the highest weight vector $[\ket{v_{max}}]$
\begin{equation}\label{cproj}
\mathcal{C}_S=\{\pi(U)[\ket{v_{max}}]:U\in K\}.
\end{equation}
Equivalently, we may consider $\mathcal{C}_S$ as a manifold $K/K_v$, where
$K_v$ is the isotropy group of $\ket{v_{max}}$,
\begin{equation}\label{Kv}
K_v=\{U\in K: \pi(U)\ket{v_{max}}=e^{i\alpha}\ket{v_{max}}, \mathrm{\ for\ some\ }
\alpha\in\mathbb{R}\}.
\end{equation}
The isotropy group of the maximum weight vector $[\ket{v_{max}}]$ depends not
only on the group $K$, but also on the representation $\pi$.

The imaginary part of the scalar product in a complex Hilbert space
$\mathcal{H}$ defines a linear symplectic structure in it (or more precisely
in the real space obtained from $\mathcal{H}$ by treating real and imaginary
parts of the components of vectors in $\mathcal{H}$ as independent real
components of vectors in a real vector space). This symplectic structure is
inherited by $\mathbb{P}\mathcal{H}$. The manifold (\ref{cproj}) treated as a
submanifold in $\mathbb{P}\mathcal{H}$ happens to be a symplectic one, ie.
can be used as a classical phase space. In addition it is a complex manifold
(in fact a unique one among the orbits of $K$ in $\mathbb{P}\mathcal{H}$),
which makes it a K\"ahler manifold \cite{guillemin84}.

The set of coherent states $\mathcal{C}_S$, ie.\ the orbit through the
highest weight vector can be uniquely characterized by certain bilinear
condition \cite{lichtenstein82}. Let us denote by $\boldsymbol\delta$ the
vector which is the half of the sum of all positive roots
$\boldsymbol\alpha$. A state $\ket{u}$ is coherent if and only if
\begin{equation}\label{licht}
\mathbf{L}\Big(\ket{u}\otimes\ket{u}\Big)=
\langle2\boldsymbol\lambda+2\boldsymbol\delta,
2\boldsymbol\lambda\rangle\Big(\ket{u}\otimes\ket{u}\Big),
\end{equation}
where $\langle,\rangle$ is the scalar product in the $r$-dimensional
Euclidean space of vectors $\boldsymbol\alpha$ and $\boldsymbol\delta$, and
\begin{equation}\label{licht2}
\mathbf{L}=\pi(C_2)\otimes I+I\otimes\pi(C_2)
+\sum_{\alpha>0}\Big(\pi(X_\alpha)\otimes\pi(X_{-\alpha})
+\pi(X_{-\alpha})\otimes\pi(X_{\alpha})\Big)+
2\sum_{i=1}^r \pi(H_i)\otimes\pi(H_i).
\end{equation}
We used the above formula, adapted to the case of $SU(2)$, to write the
bilinear characterization of spin-coherent states (\ref{klyachko}), however
the other bilinear characteristics of `classical' states (cf. (\ref{prec}),
(\ref{gen-ferm}, and (\ref{gen-bos})) where not straightforwardly derived
from (\ref{licht2}). The connection between two different bilinear
characterizations will be explained elsewhere.

All classes of pure states considered in the preceding sections can be
treated as particular instances of the general coherent states. Thus
\begin{enumerate}
\item spin $j$ coherent states are generalized coherent states for the
    $SU(2)$ group irreducibly represented in $\mathbb{C}^{2j+1}$, ie.
\begin{equation}\label{su2}
 K=SU(2),\quad \mathcal{H}=\mathbb{C}^{2j+1},
\end{equation}
\item for the fermionic separable states the appropriate identifications
    read
\begin{equation}\label{ferm}
K=SU(N), \quad \mathcal{H}=\mathcal{A}(\mathbb{C}^N\otimes\mathbb{C}^N),
\end{equation}
where $\mathcal{A}$ denotes the antisymmetric part of the tensor product,
\item for the bosonic separable states we have
\begin{equation}\label{bos}
K=SU(N), \quad \mathcal{H}=\mathcal{S}(\mathbb{C}^N\otimes\mathbb{C}^N),
\end{equation}
where $\mathcal{S}$ denotes the symmetric part of the tensor product,
\item for the separable states of two distinguishable particles,
\begin{equation}\label{qud}
K=SU(N)\times SU(M), \quad \mathcal{H}=\mathbb{C}^N\otimes\mathbb{C}^M.
\end{equation}

\end{enumerate}

\section{Low dimensional `classical' states and orthogonal symmetries}

Low-dimensional states considered above, apart from the exhibited properties
of being effectively characterizable by concurrences based on antilinear
operators, have some additional nice representations which we would like to
outline shortly in the present section.

Let us start from the following observation. A point in complex projective
space of an arbitrary finite dimension (i.e.\ a pure state of a quantum
system with an $N$-dimensional Hilbert space $\mathcal{H}$) can be always
represented in the form
\begin{equation}\label{ing}
\ket{\psi}=\cos\theta\,\mathbf{x}+i\sin\theta\,\mathbf{y},
\end{equation}
where $\mathbf{x}$ and $\mathbf{y}$ are real, unit, orthogonal vectors
\begin{equation}\label{xy}
\mathbf{x}^2=\mathbf{y}^2=1, \quad \mathbf{x}\cdot\mathbf{y}=0.
\end{equation}
Indeed, an arbitrary vector $\ket{\psi}\in\mathcal{H}$ can be decomposed into
its real and imaginary parts,
$\ket{\psi}=\mathbf{x}^\prime+i\mathbf{y}^\prime$. The normalization
condition $1=\bk{\psi}{\psi}=\mathbf{x}^{\prime 2}+\mathbf{y}^{\prime 2}$
leads to $\mathbf{x}^\prime=\cos\theta\,\mathbf{x}$, $\mathbf{y}^\prime=
\sin\theta\,\mathbf{y}$ with real, unit vectors $\mathbf{x}$ and
$\mathbf{y}$. Since we work in the projective space $\mathbb{P}\mathcal{H}$,
we still have at our disposal a phase factor, $\ket{\psi}\sim
e^{i\varphi}\ket{\psi}$, which, as simple calculation shows, is enough to
enforce orthogonality of $\mathbf{x}$ and $\mathbf{y}$. Obviously, by
alternating the global signs of $\mathbf{x}$ and/or $\mathbf{y}$, we may
restrict $\theta$ to the interval $[0,\frac{\pi}{4}]$.

The representation (\ref{ing}), (\ref{xy}) is, in general, not particularly
useful or interesting, since it is not invariant under general unitary
transformations of $\mathcal{H}$, which are allowed by quantum mechanics, but
in general do not leave $\mathbf{x}$ and $\mathbf{y}$ real. If, however, for
some reasons the relevant transformations made on the system in question
belong to the orthogonal group $SO(N)$, the situation changes.

For all low dimensional cases considered in the previous sections this is the
case. For two qubits, although the Hilbert space is four dimensional, with
the `natural' symmetry group $SU(4)$, the relevant subgroup of the local
unitary transformations which do not change the entanglement is $SU(2)\times
SU(2)$, homomorphic to the orthogonal group $SO(4)$. For two spin-3/2
fermions the Hilbert space is six dimensional but the relevant subgroup of
$SU(6)$ of transformations which respect the antisymmetry of states is
$SU(4)$ --- the full symmetry group of the four-dimensional single-particle
space (the antisymmetry is not destroyed if both particles undergo the same
operation). Here again we have a homomorphism with an orthogonal group:
$SU(4)\sim SO(6)$. Finally, in the case of two bosons in three-dimensional
space (two-dimensional single particle space) and coherent spin-1 states the
relevant subgroup of $SU(3)$ is $SU(2)$ --- the full symmetry group in the
single-particle space in the case of bosons and the group defining the
structure of coherent states, and the relevant homomorphism is the familiar
one $SU(3)\sim SO(3)$.

Throughout we consider the case when a group of transformations leaves the
set of `classical' states invariant (or more generally, when the generalized
concurrence is constant on the orbits of the group). A recurrent theme is
that the group is, on the one hand a unitary subgroup $SU(K)$ of the special
unitary group $SU(N)$ of the underlying Hilbert space, and on the other hand
it is homomorphic to the special orthogonal group $SO(N)$. The examples are
exhausted by the following table:
\begin{equation}
\begin{array}{ccccccl}
 SU(6) & \supset &       SU(4)       & \sim & SO(6) & &       \mbox{(spin-3/2
fermions),}         \\
\\
 SU(4) & \supset & SU(2)\times SU(2) & \sim & SO(4) & &
\mbox{(qubits),}          \\
\\
 SU(3) & \supset &       SU(2)       & \sim & SO(3) & &\mbox{(2-dim bosons,
spin-1 coherent states),} \\
\end{array}
\end{equation}
There are no more homomorphisms of the above type, as it is clear from the
inspection of the existing isomorphisms between unitary and orthogonal Lie
algebras (c.f.\ \cite{helgason78}, p. 519-520). This observation underscores
again the exceptional features of low-dimensional systems.

To take direct advantage of the representation (\ref{ing}) one should
determine the basis in which the relevant group is represented by real
matrices. A moment of reflection suffices to establish that such a basis
consists of vectors fulfilling $\ket{\xi_k}=\Theta\ket{\xi_k}$. Indeed, in
this basis a unitary transformation $\ket{\psi}\mapsto U\ket{\psi}$ leaving
$\bra{\psi_1}\Theta\ket{\psi_2}$ invariant, fulfills $U^TU=I$, i.e. $U$ is
orthogonal (and unitary, hence real). Such bases are dubbed `magic'---for two
qubits the magic basis consists of four orthogonal Bell states
\cite{wootters98}, in the fermionic and bosonic cases they are given in
\cite{eckert02}.

The angle $\theta$ and vectors $\mathbf{x}$ and $\mathbf{y}$ for `classical'
states must fulfil
$\bra{\psi}T\ket{\psi^\ast}=(\cos\theta\mathbf{x}-i\sin\theta\mathbf{y})\cdot
T \cdot(\cos\theta\mathbf{x}-i\sin\theta\mathbf{y})=0$. For spin-1 coherent
states this gives particularly simple condition
$\cos^2\theta-\sin^2\theta=0$, i.e. $\theta=\pi/4$.

\section{Acknowledgments}
The support by SFB/TR12 'Symmetries and Universality in Mesoscopic Systems'
program of the Deutsche Forschungsgemeischaft and Polish MNiSW grant no.\
DFG-SFB/38/2007 is gratefully acknowledged. IB was supported by the Swedish
Research Council, VR.
\section{Appendix}
We will prove that positivity of the matrix $Z$ defined by Giraud et al.\ is
equivalent to $2\,\tr(R^2)-(\tr R)^2\le 0$. For $\rho$ given by
(\ref{rhocan}) we get
\begin{equation}\label{Z}
Z=\left[
\begin {array}{ccc} -{\lambda_1}+{\lambda_2}+{\lambda_3}-4\,{{
v_1}}^{2}&-4\,{ v_2}\,{ v_1}&-4\,{ v_3}\,{ v_1}
\\\noalign{\medskip}-4\,{ v_2}\,{ v_1}&{\lambda_1}-{\lambda_2}
+{\lambda_3}-4\,{{ v_2}}^{2}&-4\,{ v_2}\,{ v_3}
\\\noalign{\medskip}-4\,{ v_3}\,{ v_1}&-4\,{ v_2}\,{ v_3}&{
\lambda_1}+{\lambda_2}-{\lambda_3}-4\,{{ v_3}}^{2}\end {array}
\right].
\end{equation}
After a short calculation we establish that the coefficients of
characteristic polynomial $P_Z(x)=x^3+a_2x^2+a_1x+a_0$ of $Z$ read
\begin{eqnarray}\label{charpolyZ}
a_2&=&4(v_1^2+v_2^2+v_3^2)-1,
\nonumber \\
a_1&=&8\det\rho+4(\lambda_1\lambda_2+\lambda_1\lambda_3+\lambda_2\lambda_3)
-8\lambda_1\lambda_2\lambda_3-1,
\nonumber \\
a_0&=&2\tr(R^2)-(\tr R)^2.
\end{eqnarray}
We will prove the following
\begin{lem}
Let: $r_i>0$, $s_i\ge 0$, $i=1,2,3$ and
\begin{eqnarray}
&&r_1+r_2+r_3=1 \\
&&r_1r_2r_3\ge r_1 s_1+r_2 s_2+r_3 s_3.
\end{eqnarray}
\end{lem}
Then
\begin{equation}
s_1+s_2+s_3<\frac{1}{4}.
\end{equation}

{\bf Proof}

Let, eg., $1>r_1\ge r_2\ge r_3 >0$. Then
\begin{equation}\label{as1}
r_1 r_2 r_3\ge r_3(s_1+s_2+s_3)+(r_1-r_3)s_1
+(r_2-r_3)s_2\ge r_3(s_1+s_2+s_3).
\end{equation}
Hence
\begin{equation}\label{as2}
r_1 r_2\ge s_1+s_2+s_3.
\end{equation}
From the arithmetic-geometric mean inequality
\begin{equation}
\sqrt{r_1r_2}\le\frac{r_1+r_2}{2}<\frac{1}{2}\ ,
\end{equation}
hence
\begin{equation}
\frac{1}{4}>r_1r_2\ge s_1+s_2+s_3.
\end{equation}
$\Box$

We substitute in the above lemma $r_i=\lambda_i$ and $s_i=v_i^2$. Then
(\ref{as1}) and (\ref{as2}) reduce, respectively, to the true statements
$\tr\rho=1$ and $\det\rho>0$. As a consequence we get $a_2<0$. We have thus
the following possibilities
\begin{enumerate}
\item $a_0<0$, $a_1>0$

In this case the signs of the coefficients of $P_Z(x)$ are $+-+-$,
whereas  those of $P_Z(-x)$ read $----$, which means that all three roots
of $P_Z(x)$ are positive, hence $Z$ is positively definite.

\item $a_0>0$, $a_1>0$

The signs of the coefficients of $P_Z(x)$ read $+-++$, i.e.\ $P_Z(x)$ has
at most two positive roots. Consequently, at least one root is negative
and $Z$ is not positively definite.

\item $a_0>0$, $a_1<0$

The signs of the coefficients of $P_Z(x)$ read $+--+$ which again gives
at most two positive roots, hence $Z$ is not positively definite.
\end{enumerate}

The remaining case $a_0<0$, $a_1<0$ is excluded. Indeed, we will prove
$a_0<0\Rightarrow a_1>0$. To this end assume as previously
$\lambda_1>\lambda_2>\lambda_3>0$ and define:
\begin{equation}
 q_1=\lambda_2+\lambda_3-\lambda_1,\quad
 q_2=\lambda_1+\lambda_3-\lambda_2,\quad
 q_3=\lambda_1+\lambda_2-\lambda_3.
\end{equation}
The implication we want to prove reduces now to
\begin{eqnarray}
 q_1 q_2 q_3&>&4 q_1 q_2v_3^2+4 q_1 q_3v_2^2+  4 q_2 q_3v_1^2 \label{z} \\
 &\Downarrow& \nonumber \nonumber \\
 q_1 q_2+ q_1 q_3+ q_2 q_3&>&
4( q_1+ q_2)v_3^2+4( q_1+ q_3)v_2^2+4( q_2+ q_3)v_1^2.
\label{t}
\end{eqnarray}
We have $ q_1+ q_2+ q_3=\lambda_1+\lambda_2+\lambda_3=1$ and $ q_2>0$, $
q_3>0$, whereas the sign of $ q_1$ can be arbitrary.

If $ q_1>0$ we have from (\ref{z})
\begin{equation}
 q_1 q_2 q_3>4 q_1 q_2v_3^2+4 q_1 q_3v_2^2+  4 q_2 q_3v_1^2>4 q_1 q_3v_2^2+
4 q_2 q_3v_1^2,
\end{equation}
hence
\begin{equation}\label{t1}
 q_1 q_2>4 q_1v_2^2+4 q_2v_1^2.
\end{equation}
Analogously
\begin{equation}\label{t2}
 q_1 q_3>4 q_1v_3^2+4 q_3v_1^2,
\end{equation}
and
\begin{equation}\label{t3}
 q_2 q_3>4 q_3v_2^2+4 q_2v_3^2.
\end{equation}
Adding (\ref{t1})-(\ref{t3}) we obtain the desired result.


\begin{thebibliography}{29}
\expandafter\ifx\csname natexlab\endcsname\relax\def\natexlab#1{#1}\fi
\expandafter\ifx\csname bibnamefont\endcsname\relax
  \def\bibnamefont#1{#1}\fi
\expandafter\ifx\csname bibfnamefont\endcsname\relax
  \def\bibfnamefont#1{#1}\fi
\expandafter\ifx\csname citenamefont\endcsname\relax
  \def\citenamefont#1{#1}\fi
\expandafter\ifx\csname url\endcsname\relax
  \def\url#1{\texttt{#1}}\fi
\expandafter\ifx\csname urlprefix\endcsname\relax\def\urlprefix{URL }\fi
\providecommand{\bibinfo}[2]{#2} \providecommand{\eprint}[2][]{\url{#2}}

\bibitem[{\citenamefont{Giraud et~al.}(2008)\citenamefont{Giraud, Braun, and
  Braun}}]{giraud08}
\bibinfo{author}{\bibfnamefont{O.}~\bibnamefont{Giraud}},
  \bibinfo{author}{\bibfnamefont{P.}~\bibnamefont{Braun}}, \bibnamefont{and}
  \bibinfo{author}{\bibfnamefont{D.}~\bibnamefont{Braun}},
  \bibinfo{journal}{Phys. Rev. A} \textbf{\bibinfo{volume}{78}},
  \bibinfo{pages}{042112} (\bibinfo{year}{2008}).

\bibitem[{\citenamefont{Perelomov}(1986)}]{perelomov86}
    \bibinfo{author}{\bibfnamefont{A.}~\bibnamefont{Perelomov}},
  \emph{\bibinfo{title}{Generalized coherent states and their applications}}
  (\bibinfo{publisher}{Springer}, \bibinfo{address}{Heidelberg},
  \bibinfo{year}{1986}).

\bibitem[{\citenamefont{Delbourgo and Fox}(1977)}]{delbourgo77a}
    \bibinfo{author}{\bibfnamefont{L.}~\bibnamefont{Delbourgo}}
    \bibnamefont{and}
  \bibinfo{author}{\bibfnamefont{J.~R.} \bibnamefont{Fox}},
  \bibinfo{journal}{J. Phys. A: Math. Gen. Phys.}
  \textbf{\bibinfo{volume}{10}}, \bibinfo{pages}{1233} (\bibinfo{year}{1977}).

\bibitem[{\citenamefont{Klyachko}(2008)}]{klyachko08}
    \bibinfo{author}{\bibfnamefont{A.}~\bibnamefont{Klyachko}},
  \bibinfo{journal}{arXiv preprint 0802.4008}  (\bibinfo{year}{2008}).

\bibitem[{\citenamefont{Klauder and Skagerstam}(1985)}]{klauder:85}
    \bibinfo{editor}{\bibfnamefont{J.~R.} \bibnamefont{Klauder}}
    \bibnamefont{and}
  \bibinfo{editor}{\bibfnamefont{B.-S.} \bibnamefont{Skagerstam}}, eds.,
  \emph{\bibinfo{title}{Coherent States. Applications in Physics and
  Mathematical Physics}} (\bibinfo{publisher}{World Scientific},
  \bibinfo{address}{Singapore}, \bibinfo{year}{1985}).

\bibitem[{\citenamefont{Klyachko}(2002)}]{klyachko02}
    \bibinfo{author}{\bibfnamefont{A.}~\bibnamefont{Klyachko}},
  \bibinfo{journal}{arXiv preprint quant-ph/0206012}  (\bibinfo{year}{2002}).

\bibitem[{\citenamefont{Barnum et~al.}(2003)\citenamefont{Barnum, Knill,
    Ortiz,
  and Viola}}]{barnum03}
\bibinfo{author}{\bibfnamefont{H.}~\bibnamefont{Barnum}},
  \bibinfo{author}{\bibfnamefont{E.}~\bibnamefont{Knill}},
  \bibinfo{author}{\bibfnamefont{G.}~\bibnamefont{Ortiz}}, \bibnamefont{and}
  \bibinfo{author}{\bibfnamefont{L.}~\bibnamefont{Viola}},
  \bibinfo{journal}{Phys. Rev. A} \textbf{\bibinfo{volume}{68}},
  \bibinfo{pages}{032308} (\bibinfo{year}{2003}).

\bibitem[{\citenamefont{Barnum et~al.}(2004)\citenamefont{Barnum, Knill,
    Ortiz,
  Somma, and Viola}}]{barnum04}
\bibinfo{author}{\bibfnamefont{H.}~\bibnamefont{Barnum}},
  \bibinfo{author}{\bibfnamefont{E.}~\bibnamefont{Knill}},
  \bibinfo{author}{\bibfnamefont{G.}~\bibnamefont{Ortiz}},
  \bibinfo{author}{\bibfnamefont{R.}~\bibnamefont{Somma}}, \bibnamefont{and}
  \bibinfo{author}{\bibfnamefont{L.}~\bibnamefont{Viola}},
  \bibinfo{journal}{Phys. Rev. Lett} \textbf{\bibinfo{volume}{92}},
  \bibinfo{pages}{107902} (\bibinfo{year}{2004}).

\bibitem[{\citenamefont{Barnum et~al.}(2005)\citenamefont{Barnum, Ortiz,
    Somma,
  and Viola}}]{barnum05}
\bibinfo{author}{\bibfnamefont{H.}~\bibnamefont{Barnum}},
  \bibinfo{author}{\bibfnamefont{G.}~\bibnamefont{Ortiz}},
  \bibinfo{author}{\bibfnamefont{R.}~\bibnamefont{Somma}}, \bibnamefont{and}
  \bibinfo{author}{\bibfnamefont{L.}~\bibnamefont{Viola}},
  \bibinfo{journal}{Int. J. Theo. Phys.} \textbf{\bibinfo{volume}{44}},
  \bibinfo{pages}{2127} (\bibinfo{year}{2005}).

\bibitem[{\citenamefont{Bengtsson and \.Zyczkowski}(2006)}]{bengtsson06}
    \bibinfo{author}{\bibfnamefont{I.}~\bibnamefont{Bengtsson}}
    \bibnamefont{and}
  \bibinfo{author}{\bibfnamefont{K.}~\bibnamefont{\.Zyczkowski}},
  \emph{\bibinfo{title}{Geometry of Quantum States}}
  (\bibinfo{publisher}{Cambridge University Press},
  \bibinfo{address}{Cambridge}, \bibinfo{year}{2006}).

\bibitem[{\citenamefont{Schliemann et~al.}(2001)\citenamefont{Schliemann,
  Cirac, Ku\'s, Lewenstein, and Loss}}]{sckll01}
\bibinfo{author}{\bibfnamefont{J.}~\bibnamefont{Schliemann}},
  \bibinfo{author}{\bibfnamefont{J.~I.} \bibnamefont{Cirac}},
  \bibinfo{author}{\bibfnamefont{M.}~\bibnamefont{Ku\'s}},
  \bibinfo{author}{\bibfnamefont{M.}~\bibnamefont{Lewenstein}},
  \bibnamefont{and} \bibinfo{author}{\bibfnamefont{D.}~\bibnamefont{Loss}},
  \bibinfo{journal}{Phys. Rev. A} \textbf{\bibinfo{volume}{64}},
  \bibinfo{pages}{022303} (\bibinfo{year}{2001}).

\bibitem[{\citenamefont{Eckert et~al.}(2002)\citenamefont{Eckert, Schliemann,
  Bru{\ss}, and Lewenstein}}]{eckert02}
\bibinfo{author}{\bibfnamefont{K.}~\bibnamefont{Eckert}},
  \bibinfo{author}{\bibfnamefont{J.}~\bibnamefont{Schliemann}},
  \bibinfo{author}{\bibfnamefont{D.}~\bibnamefont{Bru{\ss}}}, \bibnamefont{and}
  \bibinfo{author}{\bibfnamefont{M.}~\bibnamefont{Lewenstein}},
  \bibinfo{journal}{Ann. Phys.} \textbf{\bibinfo{volume}{299}},
  \bibinfo{pages}{88} (\bibinfo{year}{2002}).

\bibitem[{\citenamefont{Wootters}(1998)}]{wootters98}
    \bibinfo{author}{\bibfnamefont{W.~K.} \bibnamefont{Wootters}},
  \bibinfo{journal}{Phys. Rev. Lett.} \textbf{\bibinfo{volume}{80}},
  \bibinfo{pages}{2245} (\bibinfo{year}{1998}).

\bibitem[{\citenamefont{Uhlmann}(2000)}]{uhlmann00}
    \bibinfo{author}{\bibfnamefont{A.}~\bibnamefont{Uhlmann}},
  \bibinfo{journal}{Phys. Rev. A} \textbf{\bibinfo{volume}{62}},
  \bibinfo{pages}{032307} (\bibinfo{year}{2000}).

\bibitem[{\citenamefont{Mintert et~al.}(2005)\citenamefont{Mintert, Carvalho,
  Ku\'s, and Buchleitner}}]{mckb05}
\bibinfo{author}{\bibfnamefont{F.}~\bibnamefont{Mintert}},
  \bibinfo{author}{\bibfnamefont{A.~R.~R.} \bibnamefont{Carvalho}},
  \bibinfo{author}{\bibfnamefont{M.}~\bibnamefont{Ku\'s}}, \bibnamefont{and}
  \bibinfo{author}{\bibfnamefont{A.}~\bibnamefont{Buchleitner}},
  \bibinfo{journal}{Phys. Rep.} \textbf{\bibinfo{volume}{415}},
  \bibinfo{pages}{207–} (\bibinfo{year}{2005}).

\bibitem[{\citenamefont{Mintert et~al.}(2004)\citenamefont{Mintert, Ku\'s,
    and
  Buchleitner}}]{mkb04}
\bibinfo{author}{\bibfnamefont{F.}~\bibnamefont{Mintert}},
  \bibinfo{author}{\bibfnamefont{M.}~\bibnamefont{Ku\'s}}, \bibnamefont{and}
  \bibinfo{author}{\bibfnamefont{A.}~\bibnamefont{Buchleitner}},
  \bibinfo{journal}{Phys. Rev. Lett.} \textbf{\bibinfo{volume}{92}},
  \bibinfo{pages}{167902} (\bibinfo{year}{2004}).

\bibitem[{\citenamefont{Walborn et~al.}(2006)\citenamefont{Walborn, Ribeiro,
  Davidovich, Mintert, and Buchleitner}}]{walborn06}
\bibinfo{author}{\bibfnamefont{S.}~\bibnamefont{Walborn}},
  \bibinfo{author}{\bibfnamefont{P.~S.} \bibnamefont{Ribeiro}},
  \bibinfo{author}{\bibfnamefont{L.}~\bibnamefont{Davidovich}},
  \bibinfo{author}{\bibfnamefont{F.}~\bibnamefont{Mintert}}, \bibnamefont{and}
  \bibinfo{author}{\bibfnamefont{A.}~\bibnamefont{Buchleitner}},
  \bibinfo{journal}{Nature} \textbf{\bibinfo{volume}{440}},
  \bibinfo{pages}{1022} (\bibinfo{year}{2006}).

\bibitem[{\citenamefont{Jamio{\l}kowski}(1972)}]{jamiolkowski72}
    \bibinfo{author}{\bibfnamefont{A.}~\bibnamefont{Jamio{\l}kowski}},
  \bibinfo{journal}{Rep. Math. Phys.} \textbf{\bibinfo{volume}{3}},
  \bibinfo{pages}{275} (\bibinfo{year}{1972}).

\bibitem[{\citenamefont{Choi}(1975)}]{choi75}
    \bibinfo{author}{\bibfnamefont{M.~D.} \bibnamefont{Choi}},
  \bibinfo{journal}{Linear Alg. Appl.} \textbf{\bibinfo{volume}{10}},
  \bibinfo{pages}{285} (\bibinfo{year}{1975}).

\bibitem[{\citenamefont{Grabowski et~al.}(2006)\citenamefont{Grabowski,
    Ku\'s,
  and Marmo}}]{gkm06}
\bibinfo{author}{\bibfnamefont{J.}~\bibnamefont{Grabowski}},
  \bibinfo{author}{\bibfnamefont{M.}~\bibnamefont{Ku\'s}}, \bibnamefont{and}
  \bibinfo{author}{\bibfnamefont{G.}~\bibnamefont{Marmo}},
  \bibinfo{journal}{Open Sys. Information Dyn.} \textbf{\bibinfo{volume}{13}},
  \bibinfo{pages}{343} (\bibinfo{year}{2006}).

\bibitem[{\citenamefont{Vidal}(2000)}]{vidal00}
    \bibinfo{author}{\bibfnamefont{G.}~\bibnamefont{Vidal}},
    \bibinfo{journal}{J.
  Mod. Opt.} \textbf{\bibinfo{volume}{47}}, \bibinfo{pages}{355}
  (\bibinfo{year}{2000}).

\bibitem[{\citenamefont{Uhlmann}(1997)}]{uhlmann97}
    \bibinfo{author}{\bibfnamefont{A.}~\bibnamefont{Uhlmann}}, in
  \emph{\bibinfo{booktitle}{GROUP21, Proc. XXI Int. Coll. on Group Theoretical
  Methods in Physics, Vol.I}}, edited by \bibinfo{editor}{\bibfnamefont{H.~D.}
  \bibnamefont{Doebner}},
  \bibinfo{editor}{\bibfnamefont{P.}~\bibnamefont{Nattermann}},
  \bibnamefont{and} \bibinfo{editor}{\bibfnamefont{W.}~\bibnamefont{Scherer}}
  (\bibinfo{publisher}{World Scientific}, \bibinfo{year}{1997}), pp.
  \bibinfo{pages}{343--348}, \bibinfo{note}{quant-ph/9701014}.

\bibitem[{\citenamefont{Hughston et~al.}(1993)\citenamefont{Hughston, Jozsa,
  and Wootters}}]{hughston93}
\bibinfo{author}{\bibfnamefont{L.~P.} \bibnamefont{Hughston}},
  \bibinfo{author}{\bibfnamefont{R.}~\bibnamefont{Jozsa}}, \bibnamefont{and}
  \bibinfo{author}{\bibfnamefont{W.~K.} \bibnamefont{Wootters}},
  \bibinfo{journal}{Phys. Lett. A} \textbf{\bibinfo{volume}{183}},
  \bibinfo{pages}{14} (\bibinfo{year}{1993}).

\bibitem[{\citenamefont{Horn and Johnson}(1985)}]{horn85}
    \bibinfo{author}{\bibfnamefont{R.~A.} \bibnamefont{Horn}}
    \bibnamefont{and}
  \bibinfo{author}{\bibfnamefont{C.~R.} \bibnamefont{Johnson}},
  \emph{\bibinfo{title}{Matrix Analysis}} (\bibinfo{publisher}{Cambridge
  University Press}, \bibinfo{address}{Cambridge}, \bibinfo{year}{1985}).

\bibitem[{\citenamefont{Hall}(2003)}]{hall:03}
    \bibinfo{author}{\bibfnamefont{B.~C.} \bibnamefont{Hall}},
  \emph{\bibinfo{title}{Lie Groups, Lie Algebras, and Representations: an
  elementary introduction}} (\bibinfo{publisher}{Springer},
  \bibinfo{address}{New York}, \bibinfo{year}{2003}).

\bibitem[{\citenamefont{Guillemin and Sternberg}(1984)}]{guillemin84}
    \bibinfo{author}{\bibfnamefont{V.}~\bibnamefont{Guillemin}}
    \bibnamefont{and}
  \bibinfo{author}{\bibfnamefont{S.}~\bibnamefont{Sternberg}},
  \emph{\bibinfo{title}{Symplectic techniques in physics}}
  (\bibinfo{publisher}{Cambridge University Press},
  \bibinfo{address}{Cambridge}, \bibinfo{year}{1984}).

\bibitem[{\citenamefont{Lichtenstein}(1982)}]{lichtenstein82}
    \bibinfo{author}{\bibfnamefont{W.}~\bibnamefont{Lichtenstein}},
  \bibinfo{journal}{Proc. Am. Math. Soc.} \textbf{\bibinfo{volume}{84}},
  \bibinfo{pages}{605} (\bibinfo{year}{1982}).

\bibitem[{\citenamefont{Helagason}(1978)}]{helgason78}
    \bibinfo{author}{\bibfnamefont{S.}~\bibnamefont{Helagason}},
  \emph{\bibinfo{title}{Differential Geometry, Lie Groups, and Symmetric
  Spaces}} (\bibinfo{publisher}{Academic Press}, \bibinfo{address}{Boston},
  \bibinfo{year}{1978}).

\bibitem[{\citenamefont{Serre}(2002)}]{serre02}
    \bibinfo{author}{\bibfnamefont{D.}~\bibnamefont{Serre}},
  \emph{\bibinfo{title}{Matrices: Theory and applications}}
  (\bibinfo{publisher}{Springer}, \bibinfo{year}{2002}).

\end{thebibliography}

\end{document}